\documentclass[nohyper,12pt,letterpaper]{JHEP}
\usepackage{epsfig}

\newfont{\frak}{eufm10 scaled 1200}

\newfont{\Bbb}{msbm10 scaled 1200}     
\newcommand{\mathbb}[1]{\mbox{\Bbb #1}}
\DeclareSymbolFont{AMSa}{U}{msa}{m}{n}
\DeclareSymbolFont{AMSb}{U}{msb}{m}{n}
\let\Box\relax
\DeclareMathSymbol{\Box}{\mathord}{AMSa}{"03}

\def \eqn#1#2{\begin{equation}#2\label{#1}\end{equation}}

\title{Black Crunch}
\author{T. Banks${}^*$\\
   Department of Physics and Astronomy - NHETC\\
   Piscataway, NJ 08540\\
   and\\
   Department of Physics, SCIPP\\
   University of California, Santa Cruz, CA 95064\\
E-mail: \email{banks@physics.rutgers.edu}}

\author{W.Fischler${}^{**}$\\
Department of Physics and Astronomy\\ University of Texas, Austin,
TX
\\ E-mail: \email{fischler@physics.utexas.edu}}

\abstract{We study the growth of fluctuations in collapsing cosmologies,
extending old work of Lifschitz and Khalatnikov. As examples of systems where
the fluctuations have a different composition than the background we
study scalar fields with general improvement terms.  Fluctuations always
grow, and often dominate the homogeneous background. We argue that even for
very dilute fluctuations, scattering processes inevitably lead to a dense
gas of black holes. This leads us
to hypothesize that the generic final state of a Big Crunch is described
by a collapsing $p=\rho$ FRW cosmology. We conjecture that the black hole
fluid is invariant under the conformal Killing symmetry of this metric,
so that the final state is in fact stationary.  }

\keywords{Cosmological singularities}

\received{???????? ?st, 2000} \accepted{???????? ?th, 1998}
\preprint{\hepth{0212113}\\RUNHETC-2002-36\\SCIPP-02/26\\UTTG-13-02}

\begin{document}




\section{\bf Introduction}


A wide variety of cosmological initial conditions for Einstein's equations
lead to a future spacelike singularity colloquially known as a Big Crunch.
The energy density and curvature invariants all become singular on an 
entire spacelike
hypersurface.  For many years, physicists have speculated about the
meaning of this singularity.  A popular scenario, which has recently been
revived in the context of string theory \cite{ek} is that quantum
effects would lead to a "bounce" followed by reexpansion of the universe.

In the present paper, we will present a different view.  We will argue
that the generic final state in a Big Crunch is a maximally stiff 
$p=\rho$ fluid,
and further, that even the collapsing cosmological solution in the presence
of such a fluid, is in some sense, a stationary state.
In earlier papers \cite{bf1}\cite{bf} we have argued that such a fluid is the
appropriate semiclassical initial state for the Big Bang.
In particular, we showed there that a mechanistic model for such a fluid
is a "dense gas of black holes".  This picture will form the basis 
for our intuitive
discussion of the Big Crunch\footnote{We emphasize that although the 
equation of state is
the same, the quantum states of the Big Bang and Big Crunch are very different.
The Big Bang state consists of a collection of systems with limited 
correlation between
them because of the existence of a finite particle horizon.  In the 
Big Crunch, we imagine
that all the states of the universe are correlated.}.

One of the motivations for this picture of the Big Crunch is a series of
early papers by Lifshitz and Khalatnikov\cite{lk}(LK). These authors 
developed a general
formalism for the study of fluctuations, and behavior near a 
singularity, and applied
it primarily to the study of cosmologies with the conventional 
equations of state of
nonrelativistic and ultrarelativistic gases.  They concluded that in 
these cosmologies,
fluctuations come to dominate the energy density as one approaches 
the singularity.
We will review this calculation in Section 2.

The LK analysis applies to general equations of state of the form
$p= \kappa \rho$. They assume that the inhomogeneous fluctuations
satisfy the same equation of state.  Clearly, this is a rather
special assumption.  If we want to study a more general class of
fluctuations we must present a real field theoretical model for
them, rather than characterize them by an equation of state.  We
will do this in Section 3 by studying scalar fields with a general
coefficient of the improvement term in the stress tensor.  Again
we find growing fluctuations near a Big Crunch.  In section 4, we
discuss the fate of these fluctuations.  For the case of
fluctuations that can be thought of as a dilute gas of
relativistic particles (which we argue is a sort of worst case
scenario for our conjecture) we argue that particle collisions
will lead to a dense fluid of black holes with equation of state
$p=\rho$. We then outline our reasons for believing that this
system is conformally invariant under the conformal Killing
transformation of the $p=\rho$ F(riedmann)-R(obertson)-W(alker)
cosmology. This implies that the $p=\rho$ fluid is in some sense a
quiescent, stationary state.  All flat FRW universes whose scale
factor is a pure power of the time, have conformal Killing
vectors.  The metric rescales by a constant factor when the cosmic
time, and spatial coordinates are rescaled in a correlated
fashion.  This is in no sense a symmetry of the physics in a
generic FRW universe.  The Hamiltonian for particles and waves
propagating in such a universe are not invariant under this
conformal isometry.  However, if we believe that the $p=\rho$
fluid has no such localized excitations, then there are no
apparent observables that detect the change of scale \footnote{We
will avoid talking to any person who asks us the unpleasant
question of what the observables in a $p=\rho$ gas are!}.

The rigorous definition of black hole regions of spacetime is,
``The complement of the causal past of null infinity''.  However,
there is clearly a more local, approximate notion of black hole
associated with the creation of trapped surfaces.  We think that
a quantum version of the Cosmic Censorship conjecture would read
something like the following: {\it Any time a trapped surface forms, and
classical GR predicts a singularity, a region containing the trapped
surface is excited to a state of maximal entropy, that is a superposition
of a large number of almost degenerate states\footnote{The number is
the exponential of one quarter the area of the region,
and the typical energy splitting is of order the inverse of this number.}.
This region is then called a black hole.}  A dense fluid of thus defined
black holes, with typical separation of order their size, satisfies
$\sigma \propto \rho^{1/2}$, where $\sigma$ and $\rho$ are the entropy
and energy densities.  If, as time goes on, the black holes merge in order
to preserve this relation, then the fluid has equation of state $p=\rho$.
This is the dense black hole fluid that we discussed in \cite{bf}.

\section{\bf Review of the Lifshitz-Khalatnikov analysis}

In most of this paper we will study a flat infinite Big Crunch solution. Some
remarks about the effect of the global topology of the Big Crunch universe
will be presented in the Conclusions.  Following
\cite{lk}, we work in conformal coordinates for the FRW universe:
\eqn{coord}{ds^2 = a^2 (\tau)( - d\tau^2 + (dx^i)^2 ).}
The linearized stress tensor has the form:
\eqn{stress}{\delta T_i^k = (p+\rho) (u_i \delta u^k + u^k \delta u_i 
) + (\delta p +
\delta\rho )u_i u^k + \delta_i^k \delta p}
LK show that
\eqn{deltarho}{{\delta\rho\over\rho} = {1\over 3\rho a^2}[n^2 (\lambda + \mu) +
3{a^{\prime} \over a} \mu^{\prime}] e^{in_j x^j}.}
Primes denote derivatives with respect to conformal time $\tau$, and 
$n_i$ is the
dimensionless wave number of the fluctuation.  $\lambda$ and $\mu$ 
appear in the
parametrization of the spatial metric perturbations $h_{\alpha}^{\beta}$
\eqn{met}{h_{\alpha}^{\beta} = [\lambda (\tau) ({1\over 3} 
\delta_{\alpha}^{\beta} -
{n_{\alpha} n^{\beta} \over n^2}) + {\mu (\tau ) \over 3} 
\delta_{\alpha}^{\beta} ] e^{i
n_j x^j }.} We are in a synchronous gauge, where $h_{00} = 0 = h_{0\alpha}$.

The linearized Einstein equations determine $\lambda$ and $\mu$ in 
terms of four
auxiliary functions $\lambda_0 , \mu_0 , \psi, \zeta$.  These are 
defined by the
following four equations

\eqn{one}{\lambda_0 = - n^2 \int {d\tau \over a(\tau )}}
\eqn{two}{\mu_0 = - \lambda_0 - {3 a^{\prime} \over a^2 }}
\eqn{three}{\psi^{\prime} + \psi [{2 a^{\prime\prime} \over 
a^{\prime}} + {a^{\prime}
\over a} (-2 + 3\kappa /2)] + \kappa\zeta /2 = 0}
\eqn{four}{\zeta^{\prime} + \zeta {a^{\prime} \over a} (1 + 3\kappa 
/2 ) + \psi [- 2 n^2
+ 3 {a^{\prime\prime} \over a} + ({a^{\prime} \over a})^2 (9\kappa /2 
- 6)] = 0}
In these equations, $\kappa$ parametrizes the equation of state of both the
background and the fluctuations:
$\delta p = \kappa\delta\rho$.
Then
\eqn{five}{\lambda + \mu  = (\lambda_0 + \mu_0 )\int \psi d\tau}
\eqn{six}{\lambda^{\prime} - \mu^{\prime} = (\lambda_0^{\prime} - 
\mu_0^{\prime} ) \int
\psi d\tau + {\zeta \over a}}

For generic values of $\kappa$,
we can solve equation (2.7) for $\zeta$ and obtain an equation 
depending only on
$\psi$. This strategy does not work for $\kappa =0$, and we will 
discuss this special
case first.  The relevant formulae for the coefficients in our equations, when
the background equation of state is $p=\kappa\rho$ are:
\eqn{scala}{a = \tau^{{2\over 3\kappa +1}}}
\eqn{scalb}{{a^{\prime\prime}\over a^{\prime}} = {1 - 3\kappa \over 
(1 + 3\kappa )\tau}}
\eqn{scalc}{{a^{\prime}\over a} = {2 \over (1 + 3 \kappa) \tau}}
\eqn{scald}{{a^{\prime\prime}\over a} = {2 (1 - 3 \kappa) \over (1 + 
3 \kappa)^2 \tau^2}}
\eqn{scale}{\rho a^2 \sim \tau^{-2}}

The equation for $\psi$ when $\kappa = 0$ has the simple solution
\eqn{psisol}{\psi = \psi_0 \tau^2}
It is then easy to see that the term proportional to $\psi$ in the 
equation for $\zeta$
is negligible near the singularity.  Note that this includes all the 
dependence on the
comoving wave number, as long as that quantity is fixed.  The 
equation for $\zeta$ now
becomes

\eqn{zeta0}{\zeta^{\prime} + {2 \over  \tau} \zeta = 0.}
So
\eqn{zetasoln}{\zeta = \zeta_0 {\tau}^{ - 2 }.}

These equations can now be plugged into (2.9) and (2.10) to obtain 
$\lambda$ and $\mu$
and eventually ${\delta\rho \over \rho}$.  The lower limit of 
integration can be changed
at will by gauge transformations that preserve the synchronous gauge 
\cite{lk}.  We take
it to be zero, to cancel off singular, pure gauge terms.  Note that 
the powers of $\tau$
appearing in these terms are different than the powers in the gauge 
invariant terms that
we retain.  The first, momentum dependent, term in Equation (2.3) is 
subleading, and we
find

\eqn{delta0}{{\delta\rho \over\rho} \sim \zeta_0 \tau^{- 3}}
Inhomogeneous nonrelativistic fluctuations dominate
the background near the singularity.

We now turn to the general case, eliminating $\zeta$ in terms of 
$\psi$. We find a second
order equation

\eqn{csi}{\psi^{\prime\prime}  - 2\psi
{(1 - 3\kappa ) \over (1 + 3\kappa)^2
\tau^2 } = 0.}
This has solutions of the form $\tau^A$ where
\eqn{A}{A^2 - A - 2 { (1-3\kappa)  \over (1 + 3 \kappa
)^2}  = 0.} As before, we choose the most singular power.  The power 
law behavior of
$\zeta$ is
$\tau^{(A - 1)}$. We again find that the momentum dependence drops in 
the most singular
terms ({\it i.e.} all comoving momentum modes grow in the same way) 
and we find:
\eqn{final}{{\delta\rho\over\rho} \sim [{18 (1 + \kappa) \over 
(A+1)(1 + 3 \kappa )^2} -
{3\over 1 + 3 \kappa} + {1\over\kappa} (A - {  3\kappa + 2 \over 1 + 3\kappa})]
\tau^{x \over 1 + 3\kappa},}
where $x = A (1 + 3 \kappa ) - 2$. $x$ satisfies
\eqn{x}{x^2 + 3 x (1 - \kappa  )  = 0.}

One root of this equation vanishes, reflecting a perturbation which is simply a
rescaling of the background, while the other is negative, reflecting 
an unstable
growth of inhomogeneous fluctuations on all length scales.

\section{\bf Improved fluctuations}

The LK analysis relies on equal coefficients in the background and fluctuation
equations of state, and is inconsistent if this assumption is not 
made.  Nonetheless
it is clear that fluctuations which are not the same kind of matter 
as the homogeneous
background are possible.  In any such case, one must specify the dynamics of
the matter in a way that goes beyond the use of an equation of state. 
One could for
example consider a general scalar field Lagrangian or multiple 
scalars. It turns out
that potential terms in the Lagrangian are irrelevant near the 
singularity.  Scalars
with non-canonical kinetic terms (a metric on field space) all behave 
like a minimally
coupled scalar. Couplings of the form $f(\phi) R$, {\it do} affect 
the behavior near
the singularity.  At the linearized level in which we will be 
working, it is sufficient
to keep the improvement term\cite{ccj}.  These are the simplest models in which
we can analyze fluctuations that do not satisfy the equation of state 
of the background.

We will set the classical value of the scalar field to zero.  This 
has the advantage
that the gravitational backreaction to the fluctuations is a higher 
order correction.
Thus, we can work in FRW coordinates and many of the gauge fixing 
questions in the LK
analysis do not arise.  We are describing inhomogeneous scalar fields 
on a fixed
homogeneous background manifold, which is undergoing a Big Crunch.

The equations for scalar field perturbations in FRW coordinates are
\eqn{scaleq}{{\ddot{\phi_k}} + 3{\dot{a}\over a}\dot{\phi_k} + 
({k^2 \over a^2})
\phi_k  + \xi R\phi_k = 0.} The value $\xi = 1/6$ corresponds to 
conformal coupling.  In
this case, there is a cancellation  of the most singular term in the
fluctuations.  We will discuss it separately below.

In the mostly plus metric convention, the scalar curvature of an FRW
background satisfies
\eqn{R}{R = \rho - 3p = 3(1- 3\kappa )({\dot{a} \over a})^2 }

The scale factor $a$ satisfies $a \sim t^{\beta}$, where $\beta = {2\over
{3(1+\kappa)}}$.  The scalar equation then has solutions
of the form $\phi_k \sim t^{\alpha}$ near $t=0$.  Note that for $\kappa > -
1/3$, $\alpha$ is independent of $k$, because the momentum term in 
equation \ref{scaleq}
is subleading.  We will not study these very negative values of 
$\kappa$ because
we find it implausible that such negative pressure matter will dominate the
universe near the Crunch.  Other forms of energy density grow much more rapidly
near the singularity.  The exponent $\alpha$ satisfies
\eqn{alpheq}{\alpha^2 + (3\beta -1)\alpha + 3\xi(1 - 3\kappa)\beta^2 = 0}
whose solutions are
\eqn{alpheqsoln}{\alpha = {1\over {2 + 2\kappa}}\Bigl{[}(\kappa - 1) 
\pm \sqrt{(\kappa -
1)^2 +{16\xi \over 3}(3\kappa -1)}\Bigr{]}.}

For $\kappa > -{1\over 3}$ the argument of the square root is positive
for $0<\xi < {1\over 6}$.  It vanishes in the case $\kappa =1$ and $\xi = 0$.
In this case we have a logarithmic solution as well.
For general $\xi$ the leading singularity in the stress tensor does not
cancel and its order can be estimated just by calculating 
$\dot{\phi}^2$. 
In any FRW metric with power law scale factor, the Friedmann equation 
implies that
the background energy density scales as $t^{-2}$ near the 
singularity. Thus the power
law in ${\delta\rho \over \rho} $ is just $t^{2\alpha}$  .  For all values of
$\kappa > -{1\over 3}$ and $0 < \xi \leq {1/6}$, there is always one 
negative root
  of
$\alpha$ and we find that the fluctuations are growing relative to 
the background.
For the special case $\xi =0 $, $\kappa = 1$, both roots vanish.  This is the
situation where we also have a logarithmic solution, but this also 
gives a constant
${\delta\rho \over \rho}$.  Note however that this case is also covered by the
LK analysis, which takes into account the back reaction of the fluctuating
energy density on the metric.  That analysis seems to give a 
logarithmic singularity
in ${\delta\rho \over \rho}$.  There has been some controversy about this in
the literature\cite{contro}.

Finally, we turn to the conformally coupled case.  In this case the 
trace of the
fluctuating stress tensor vanishes identically as a consequence of the
scalar field equation of motion.  Covariant conservation then implies that
$\delta\rho \sim a^{-4} \sim t^{-{8\over {3(1+\kappa)}}}$. This grows more
rapidly than the background only when $\kappa < {1\over 3}$. Thus the 
conformally
coupled case has tamer fluctuations than more general values of $\xi$.  One
can check that this is due to a cancellation of the most singular term in the
stress tensor, for this value of $\xi$.  Again, for $\kappa = {1\over 
3}$ the LK
analysis applies and implies that gravitational backreaction makes 
the fluctuations
more singular. Intuitively this behavior is expected because gravity 
is attractive
and leads to the growth of inhomogeneous fluctuations.

In an expanding universe, finding that $\delta\rho$ grows more
rapidly than $\rho$ is a signal that one is about to enter an era
of local gravitational collapse and black hole formation.  This
would seem even more likely in the contracting case.  We can see
that the situation of a very dilute gas of conformally invariant
fluctuations is in some sense the worst case scenario for black
hole formation. In a generic background, in the approximation we
have used in this section, the conformally coupled scalar fluctuations
grow less rapidly than any other form of scalar energy density.
Similar behavior is found for photons or any other form of conformal
matter.
In the next section we will examine this case and
argue that even here, scattering processes, which are not taken
into account either in the analysis of this section, or that of
LK, will lead to black hole formation.

\section{\bf Scattering of inhomogeneities}

Let us imagine then that we begin with an FRW universe which is heading towards
a Big Crunch.  We superpose on this background a spectrum of 
fluctuations, consisting
of relativistic particles.  The energy density in the $k$th
mode at some initial time is equal to
$\epsilon_k$.  Here $k$ is the dimensionless comoving momentum. We imagine that
the particles have initial $k$ values which are all of the same order 
of magnitude.
At later times, the fluctuating energy density is
\eqn{fluqen}{\delta\rho_k \sim {\epsilon_k \over a^4}}
The typical center of mass energy in collisions of these particles 
will be $\sim k/a$.
Their number density is $n_k \sim  \epsilon_k /ka^3$.  Thus their 
typical impact
parameter is
\eqn{imp}{I \sim a {{k^{1/3}}\over \epsilon_k^{1/3}}}

There are strong arguments\cite{strong} that when the impact parameter is
of order the Schwarzschild radius of the center of mass energy, a 
finite fraction
of the collisions will lead to black hole formation.  The criterion for this to
occur is
\eqn{crit}{ a {{k^{1/3}}\over \epsilon_k^{1/3}} \sim {k\over a M_P^2}.}
This criterion is reached when
\eqn{crita}{ a^2 M_P^2 \sim \epsilon_k^{-2/3} k^{4/3}}

If $\epsilon_k / k^4 \ll 1$ then this occurs when the energy density 
of the fluctuations
is much smaller than the Planck scale.  This distribution cannot be 
thermal, but by
assumption we are talking about small inhomogeneous fluctuations on a 
homogeneous
background, so we do not expect them to be thermal.  For a given 
initial distribution
$\epsilon_k$, black hole formation will occur first at some particular value of
$k$.  Other particles in the distribution will be swallowed by the gas of black
holes that has been created.

At the time of black hole formation, the separation between black holes is
of the order of their Schwarzschild radii.  Thus, the black hole gas
satisfies the entropy density/energy density relation $\sigma \propto 
\sqrt{\rho}$.
The black holes are being pushed together by the contraction of the universe.
Thus they will begin to undergo a process of continual merger, creating the
dense black hole fluid that we have identified as a mechanistic model of
the $\rho = p$ equation of state.  Eventually the entropy density will hit the
holographic bound\cite{fsb}.  We have
argued\cite{bf} that in this situation there can be no inhomogeneous energy
fluctuations of the black hole fluid.  The entropy of the system 
resides in internal
black hole states, which are (with exponential accuracy as the black 
hole mass gets
large) degenerate in energy.  Note that this conclusion depends on 
quantum mechanics
through the Bekenstein-FSB bound.  We cannot expect to see it coming out of the
classical GR equations for fluctuations in the $p=\rho$ background. 
In particular,
although these equations can encode the equation of state, they have no
information about whether the holographic bound is saturated.  In a relatively
dilute system of black holes ({\it e.g} one in which the typical separation
between black holes is 100 times their Schwarzschild radius), the holographic
bound is not saturated.  Inhomogeneous fluctuations will obviously grow
in such a gas, resulting in the production of larger black holes.  Our point
is that the holographic bound tells us that this process eventually stops.

Although the $p=\rho$ FRW Big Crunch proceeds to a singularity in finite time,
we believe that this is an illusion.  The maximally saturated black hole fluid
seems like a quiescent stationary state.  It is self similar.  We 
conjecture that
the mathematical representation of this intuition is the invariance of the
quantum mechanics of the $p=\rho$ fluid under the conformal Killing vector
of the FRW geometry.  Indeed, a system invariant under this transformation
feels the Big Crunch as a geometry completely equivalent to flat Minkowski
space.  The singularity has disappeared because the physical system which
experiences it is invariant under the time evolution in a collapsing
FRW universe.

\section{\bf Conclusions}

In this paper, we have reviewed and extended earlier work, which shows
that fluctuations grow in Big Crunch spacetimes.  We argued that
even mild growth of fluctuations ({\it i.e.} even those for which
${\delta\rho\over\rho}$ remains bounded) would, via scattering, lead
to production of a dense black hole fluid filling spacetime.
We conjectured that this is the proper description of the final
state of a Big Crunch.  Its global geometry is that of a collapsing,
homogeneous $p=\rho$ FRW cosmology. We conjectured that the black hole
fluid was in fact invariant under the conformal Killing symmetry
of the geometry, so that the Black Crunch is in fact a stationary
state and the singularity is an illusion.  There are no observables
of the physical system filling the universe, which can feel the
singularity.

This paper dealt only with infinite flat universes.  The neglect
of spatial curvature is likely to be unimportant, but the neglect
of spatial topology might not be.
In previous work with Motl\cite{bfm} we have argued that certain
Big Crunch singularities in cosmological solutions of the low
energy field equations of M-theory might be reinterpreted by
duality transformations and the singularity resolved.  These all
have the following character: after the duality transformation the
spacetime is approaching a large radius compact geometry in either
11 dimensions, or weakly coupled type II string theory.   The
singularity consists of the fact that one approaches infinity in
moduli space in finite time.   We conjectured that production of
the light Kaluza-Klein modes of the expanding background would
slow the expansion rate and eliminate the singularity.

For such singularities, there are two competing processes going on.
Imagine approaching the singularity in a duality frame where it
looks like some of the dimensions are contracting, and let the universe
be filled primarily with unwound matter states in this frame.
The matter density is homogeneous, with small inhomogeneous fluctuations.
As we approach the singularity, scattering processes among these
states can produce light winding states, which are the light KK modes
of the nonsingular duality frame.   The inhomogeneities can also
scatter to produce black holes.  The question of which of these processes
dominate may depend on the precise initial state, and the fate of
the singularity may depend on which process dominates.  The issue is
also complicated by the possibility of black hole to black brane transitions.
So it seems possible that the mechanism of \cite{bfm} provides an
alternate endpoint for this kind of Big Crunch.  However, much work
remains to be done to sort out what really happens.

In \cite{bfm} we also pointed out another class of singularities that
could not be removed by duality transformation.  The system approaches
asymptotic regions of moduli space that are not dual to any semiclassical
description.  In this case, we expect that the Black Crunch described
in this paper will be the appropriate end point for the evolution of
the system.

There are several directions of research which are affected if the
Black Crunch is indeed the fate of cosmological singularities.  The
first is research into models where the  universe passes through
a Big Crunch and bounces back to a large expanding universe.   In our
opinion, models of that type only make sense if the singularity is
in fact dual to an expanding universe.  The analysis of what happens
as one passes through the classical singularity is very complicated
and involves the competition of black hole and KK mode production.
It seems unlikely that any straightforward translation of information
about fluctuations before the crunch into predictions about fluctuations
after the crunch could be performed.  One would have to solve a complicated
system of space time dependent rate equations.  The geometry after the crunch
is not related to that prior to the crunch by any local transformation,
and the nature of the light states changes completely.  None of the models
in the literature exhibit any of these features.

The second direction which we find problematic is the attempt to find
a resolution of singularities within the domain of weakly coupled string
theory.  In a Big Crunch singularity, particle energies are blueshifting
and impact parameters are getting smaller.  It is well known that string
perturbation theory breaks down at high energy and a range of
impact parameters that grows with the energy. The kinematic regime
that particles experience in a Big Crunch is one in which black hole
production is important.  Note that even if our fantasies about the
resolution of certain singularities by duality turn out to be correct,
the system always passes through regions where the string coupling is
of order one and string perturbation theory is an inadequate guide to
the physics.   It is Planck scale, rather than string scale physics
which is involved when it comes to the crunch.

\section{Acknowledgments}

The research of T.B was supported in part by DOE grant number 
DE-FG03-92ER40689, the
research of W.F. was supported in part by NSF grant- 0071512.

%


\end{document}